\documentclass[a4paper,11pt]{article}
\usepackage[english]{babel}
\usepackage{graphicx}

\begin{document}

\title{Gravitational lensing by gravitational waves}

\author{G. S. Bisnovatyi-Kogan$^{1,2,3}$ and O. Yu. Tsupko$^{1,3}$
\\[5mm]
$^1$Space Research Institute of Russian Academy of Science,\\
Profsoyuznaya 84/32, Moscow 117997\\
$^2$Joint Institute for Nuclear Research, Dubna, Russia\\
$^3$Moscow Engineering Physics Institute, Moscow, Russia\\ \\
\it e-mail: gkogan@iki.rssi.ru, tsupko@iki.rssi.ru\\
}

\date{}
\maketitle \abstract{Gravitational lensing by gravitational wave
is considered. We notice that although final and initial direction
of photons coincide, displacement between final and initial
trajectories occurs. This displacement is calculated analytically
for the plane gravitational wave pulse. Estimations for
observations are discussed.}

\section{Introduction}
According to general relativity, any gravitational field can
change trajectory of photons or, in other words, deflect light
rays. Hence the gravitational field may act as a gravitational
lens.

Gravitational lensing by gravitational waves in different cases
was considered by many authors (see \cite{Nov90}, \cite{Far1992},
\cite{DamEsp} and references therein). It was found that the
deflection angle vanishes for any localized gravitational wave
packet because of transversality of gravitational waves
\cite{DamEsp}. Thus if the photon passes through finite
gravitational wave pulse its deflection due to this wave is equal
to zero. Nevertheless we notice that the displacement between
trajectories of the photon before and after passing the wave may
occur.

In this work we confirm analytically vanishing of deflection angle
for plane wave pulses. However, we have found that the
gravitational wave (GW) changes the photon propagation in another
way, simply shifting its whole trajectory after passing through
the GW (see figs 1,2). This displacement is found analytically for
the photon passing through the plane GW. On the basis of this
result we obtain an approximate formula for the estimation of
observational effects.

\section{The deflection angle}

The wave vector of the photon $k^i$ (tangent to a trajectory) by
definition is equal to $k^i = \frac{dx^i}{d\lambda},$ where
$\lambda$ is a parameter changing along the trajectory \cite{LL}.
The geodesic equation is written as $Dk^i = 0$ or $Dk_i = 0$,
where $D$ denotes a covariant derivative. It is more convenient to
use the second expression. After some transformation


\begin{equation}
\left( \frac{\partial k_i}{\partial x^l} - \Gamma^k_{il} \, k_k
\right) dx^l = 0 \, , \quad \frac{\partial k_i}{\partial x^l}
\frac{d x^l}{d \lambda} =  \Gamma^k_{il} \, k_k \frac{dx^l}{d
\lambda} \, ,
\end{equation}

\begin{equation}
\frac{d k_i}{d \lambda} = \Gamma_{k, il} \, k^k k^l \, , \quad
\frac{d k_i}{d \lambda} = \frac{1}{2} \left( \frac{\partial
g_{ki}}{\partial x^l} + \frac{\partial g_{kl}}{\partial x^i} -
\frac{\partial g_{il}}{\partial x^k} \right) k^k k^l \, ,
\end{equation}
we obtain the equation of motion for the photon:

\begin{equation}  \label{eq-motion}
\frac{d k_i}{d \lambda} = \frac{1}{2} \, k^k k^l \, \frac{\partial
g_{kl}}{\partial x^i} \, , \quad \mbox{or  } \; \ddot{x}_i=
\frac{1}{2} \, \dot{x}^k \dot{x}^l \, \frac{\partial
g_{kl}}{\partial x^i},
\end{equation}
where dot denotes derivative with respect to parameter $\lambda$.

Let us consider the gravitational wave in a flat space with a
metric $g_{ik} = \eta_{ik} + h_{ik}$, $h_{ik} \ll 1$, where
$\eta_{ik}$ is a flat metric $(-1,1,1,1)$ and $h_{ik}$ is a small
perturbation (gravitational wave). In this approximation one can
integrate equation (\ref{eq-motion}), calculating right-hand side
of equation with unperturbed trajectory of the photon. Performing
integration, we obtain the expression for the deflection angle
(compare with \cite{DamEsp}):

\begin{equation} \label{Damour}
\hat{\alpha}_i = \frac{k_i(+ \infty) - k_i(- \infty)}{k} = \int
\limits_{- \infty}^{+ \infty} \frac{1}{2} \, k^k k^l \,
\frac{\partial h_{kl}}{\partial x^i} d \lambda
\end{equation}
where $h_{ik}$ is calculated along the straight line trajectory
and $k^i = const$ along unperturbed trajectory.

Consider the photon moving along $z$-axis. Its unperturbed
trajectory is $z = z_0 + c t$, and we can use the coordinate $z$
as the parameter $\lambda$. Then the wave vector is
$k^i=(1,0,0,1)$, $k=1$. When the photon passes through the finite
wave packet, we denote the $z$-coordinate of the input of the
photon into the wave front as $z_1$ and the $z$-coordinate of the
output from the wave front as $z_2$ ($z_1<z_2$). Hence we have the
expression for the deflection angle in the form (compare with
\cite{Far1992}):

\begin{equation} \label{Faraoni}
\hat{\alpha}_i =  \frac{1}{2} \int \limits_{z_1}^{z_2}
\frac{\partial}{\partial x^i} (h_{00}+2h_{03}+h_{33}) \, dz \, .
\end{equation}

Let us calculate photon deflection by the plane GW pulse. Let us
consider a light ray propagating under the angle $\varphi = - (\pi
- \theta)$ relative to the direction of the plane GW packet
propagation (see Fig.1). Let us define for convenience two
reference systems $K$ and $K'$. The first one is connected with
direction of the light ray: the photon moves along $z$-axis in a
positive direction in the reference frame $K$. The second one is
connected with the direction of propagation of the gravitational
wave pulse. The gravitational wave packet moves along $z'$-axis in
a positive direction in the reference frame $K'$ (see Fig.1). The
systems are at the rest relative to each other and their origins
of coordinates coincide. At the initial time $t=0$ the photon is
situated at $z_0<0$ $(x_0=y_0=0)$, the wave vector of the photon
is $k^i=(k^0,0,0,k^z)=(1,0,0,1)$, $k=k^z$. The form of the wave
pulse is sinusoidal (the top part of the sinusoid, with the phase
changing from $0$ to $\pi$, and with zero perturbations on the
boundaries):

\begin{equation}
h_{ik}' \propto \sin \xi', \quad \xi'=0..\pi, \quad \xi'=\omega t
- k'_g z', \quad k'_g = \omega/c,
\end{equation}

where $\omega$ and $k'_g$ are the frequency and the wave vector of
gravitational wave in $K'$ correspondingly. The pulse width
$\delta$ (in space) is $\delta=c \pi/\omega$.

It is convenient to use non-dimensional variables for time
$\tilde{t} = t/t_0$, $t_0 = 1/\omega$ and distances $\tilde{x} =
x/x_0$, $x_0 = c/\omega$. Hereafter we omit tildes for simplicity.
In non-dimensional variables the equation of motion
(\ref{eq-motion}) looks the same, and the gravitational wave form
is written as $\sin(t - z')$.

The right side of (\ref{Faraoni}) includes components $h_{00}$,
$h_{03}$, $h_{33}$, which are components of gravitational
perturbation in the reference system $K$. Gravitational wave moves
along the axis $Oz'$ in the reference system $K'$, therefore the
GW has non-zero components $h'_{11}$, $h'_{12}$, $h'_{21}$,
$h'_{22}$ only. The reference system $K$ transforms into the
system $K'$ by rotation by the angle $\varphi = - (\pi - \theta)$
around the axis $x$ (see Fig.1). Hence we have:
\begin{equation}
h_{00}=h_{03}=0 \, , \quad h_{33} = \sin^2 \varphi \, h'_{22} .
\end{equation}
Writing $h'_{22}$ as $h'_{22} =  h \sin (t - z') $, where $h$ is
the amplitude of wave, we obtain:

\begin{equation}
h_{33} = h \, \sin^2 \theta \, \sin (t + z \cos \theta - y \sin
\theta) .
\end{equation}

Taking into account that the straight line ray has the trajectory
$z = z_0 + t$, one can find the points of intersection of the
photon and the wave front. The point of the input is $z_1$, the
point of the output is $z_2$, the point of the perturbation
maximum is $z_m$ (in the reference system $K$ the gravitational
wave moves in the negative direction of the axis $z$, therefore we
have $z_1 < z_m < z_2$):
\begin{equation}
z_1 = \frac{ z_0}{1 + \cos \theta} \, , \quad z_2 = \frac{\pi +
z_0}{1 + \cos \theta} \, , \quad z_m = \frac{\pi/2 + z_0}{1 + \cos
\theta} \, .
\end{equation}

Because of the symmetry, the deflection may happen only in the
plane $(zy)$: $\alpha_y$. Let us define $F_y(z)$ as
\begin{equation}
F_y(z) = \frac{1}{2} \left.  h \, \sin^2 \theta \,\left[
\frac{\partial}{\partial y} \left( \sin (z - z_0 + z \cos \theta -
y \sin \theta) \right) \right] \right|_{y=0} = \frac{\partial
\varphi_y}{\partial y} ,
\end{equation}
$$
\varphi_y = \frac{1}{2} h \sin^2 \theta \sin (z - z_0 + z \cos
\theta - y \sin \theta),
$$
$\varphi_y = 0$ outside the GW pulse.

Then the deflection angle in the first part of the way within the
wave is:
\begin{equation}
\alpha_1 = \int \limits_{z_1}^{z_m} F_y(z) dz = - \frac{1}{2} \, h
\, \frac{\sin^3 \theta}{1 + \cos \theta} = - \frac{1}{2} \, h \,
(1 - \cos \theta) \sin \theta \, .
\end{equation}
The deflection angle in the second part of the way within the wave
is:
\begin{equation}
\alpha_2 = \int \limits_{z_m}^{z_2} F_y(z) dz =  \frac{1}{2} \, h
\, \frac{\sin^3 \theta}{1 + \cos \theta} =  \frac{1}{2} \, h \, (1
- \cos \theta) \sin \theta \, .
\end{equation}
And the total deflection angle is:
\begin{equation}
\hat{\alpha} = \int \limits_{z_1}^{z_2} F_y(z) dz =  0 \, .
\end{equation}

The top part of the sinusoid is symmetrical relative to the
vertical axis. We also have considered a non-symmetrical plane
waveform and have obtained numerically that vanishing of the
deflection angle occurs in this case too. We also checked
vanishing for different velocities of the photon (in the medium,
where photon has velocity $< c$).

\section{The displacement}

Let us find a displacement analytically for the plane
gravitational wave pulse. To find the displacement we need to
integrate farther the equation of motion (\ref{Faraoni}), what
gives:
\begin{equation}
y(z) = \int \limits_{z_1}^{z} \left[ \int \limits_{z_1}^{z'}
F_y(z") dz" \right] dz' \, , \quad y(z_1)=0 \, .
\end{equation}
We obtain for the trajectory $y(z)$ of the photon within the wave
($z_1<z<z_2$):
\begin{equation} \label{traj}
y(z) =  \frac{1}{2} \, h \, \frac {( -1 + \cos(z-z_0+z \cos\theta)
) \sin^3\theta}{(1  + \cos \theta) ^{2}} \, , \quad y(z_1)=0 \, .
\end{equation}
We see that the photon trajectory has a sinusoid form within wave.
Using (\ref{traj}), we see that the total displacement along the
axis $y$ does not vanish, and it is equal to:
\begin{equation}
\Delta y = y(z_2) - y(z_1) = - h \, \frac{\sin^3 \theta}{(1 + \cos
\theta)^2} = - h \, \frac{ 1 - \cos \theta }{1 + \cos \theta} \sin
\theta \, .
\end{equation}

Thus although initial and final directions of photon coincide, and
the deflection angle vanishes, the displacement in the trajectory
occurs. This displacement is absent in case of $\theta=0$ (the
photon and the gravitational wave directions are parallel) and
reaches its maximum in the case at $\theta=\pi/2$ (the photon and
the gravitational wave directions are orthogonal). It is clear
that this displacement will be equal to zero if we consider the
whole sinusoid with the top and the bottom parts, because the
displacement due to the top part of the sinusoid will be cancelled
by the displacement due to the bottom part. Therefore this
displacement takes place mainly in the case of isolated wave
pulses, which have a form similar to the top part of the sinusoid
or when it has the non-symmetrical top and bottom parts of wave
profile, and may, in principal, vanish for the periodic wave of a
long duration. The wave pulses may be produced, for example,
during stellar collapse (see \cite{Shapiro}, paper contains many
figures with waveforms) or during formation of large scale
structure of the Universe (see \cite{BK2004}).

We calculate displacement in the non-dimensional variables. In
dimensional variables we have:
\begin{equation}
\Delta y = - h \, \frac{\delta}{\pi}\frac{\sin^3 \theta}{(1 + \cos
\theta)^2} \, .
\end{equation}
Let us estimate observational effects caused by this displacement.

\section{The observational effects of displacement}

Directions of photons passing through the gravitational wave
packet does not change, therefore any focusing of rays does not
occur in this case. Thus the displacement in trajectories does not
lead to any magnification effect. But the displacement leads to
change of the angular position of object for distant observer. The
change of the angular position due to passing of the light ray
through the gravitational wave pulse $\Delta \alpha_d$ is as (see
fig.3)
\begin{equation}
\Delta \alpha_d = \frac{\Delta y}{D_s} \simeq \frac{h \delta}{D_s}
\, ,
\end{equation}
where $h$ is the amplitude of the GW pulse, $\delta$ is its
thickness and $D_s$ is a distance between the source and the
observer.

Let us estimate the change of the angular position for the GW
pulses produced during formation of large scale structure of the
Universe in dark matter (see \cite{BK2004}). For estimates we put
$h = 10^{-11}$, $\delta = Mpc$, $D_s = 100 Mpc$, then we obtain
\begin{equation}
\Delta \alpha_d \simeq 2 \cdot 10^{-8} \, arcsec .
\end{equation}

\section*{Acknowledgments}
Authors are thankful to M. Barkov for useful discussions. This
work was partly supported by RFBR grants 08-02-00491 and
08-02-90106, the RAN Program “Formation and evolution of stars and
galaxies” and Grant for Leading Scienti.c Schools NSh-2977.2008.2.
The work of O.Yu. Tsupko was also partly supported by the Dynasty
Foundation.

\begin{figure}
\centerline{\hbox{\includegraphics[width=1.0\textwidth]{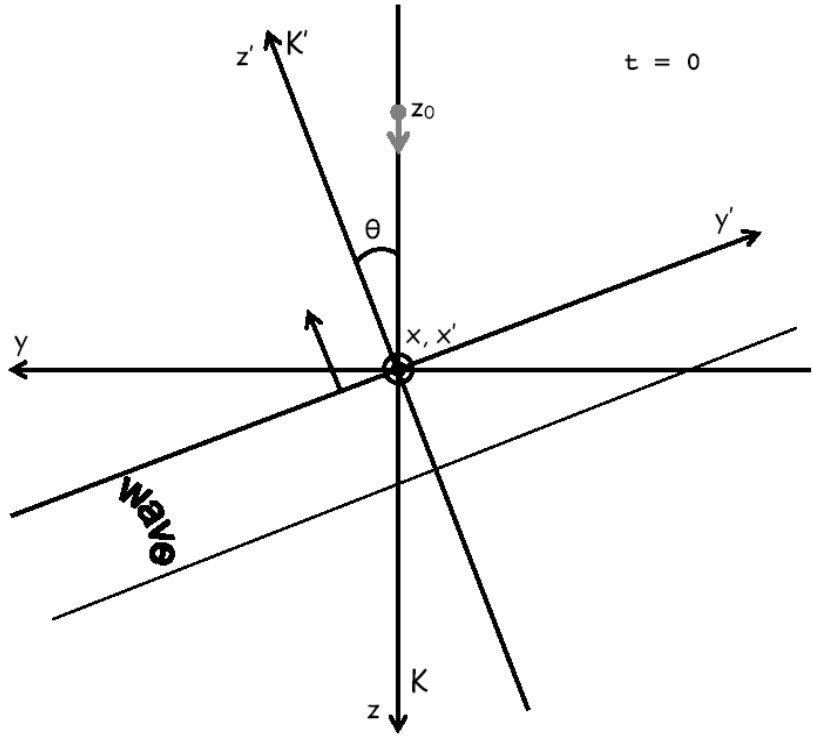}}}
\caption{Lensing of the photon by the plane GW pulse. The initial
state. The photon moves along $z$-axis  in the reference system
$K$. Gravitational wave packet moves along $z'$-axis  in the
reference system $K'$. The reference system $K$ transforms into
the system $K'$ by rotation by the angle $\varphi = - (\pi -
\theta)$ around the axis $x$ (positive rotation is anticlockwise).
}
\end{figure}

\begin{figure}
\centerline{\hbox{\includegraphics[width=1.0\textwidth]{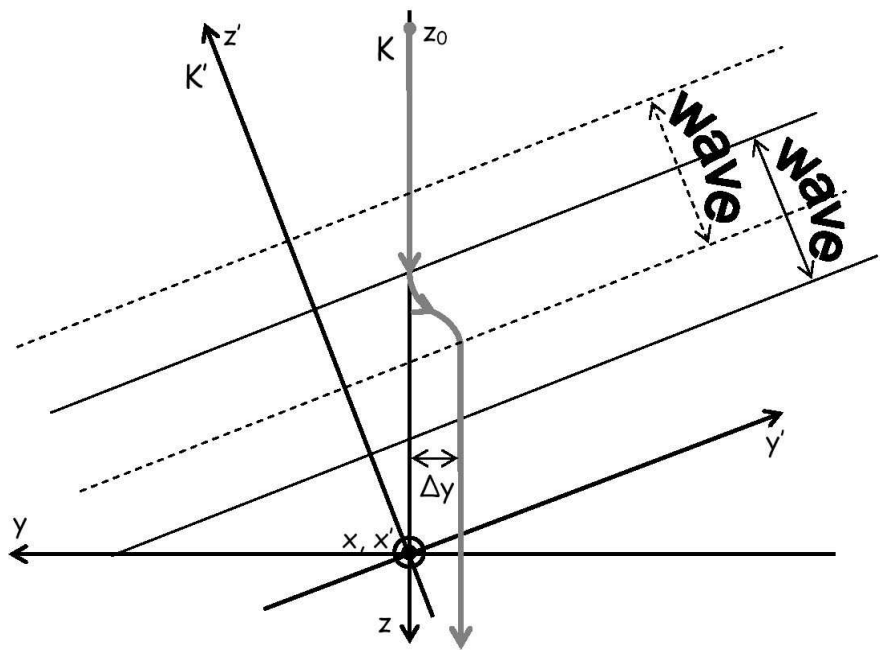}}}
\caption{Passing of the photon through the GW. The positions of
the wave packet at the time of the photon input and the photon
output are shown by full line and dashed line correspondingly.}
\end{figure}

\begin{figure}
\centerline{\hbox{\includegraphics[width=0.8\textwidth]{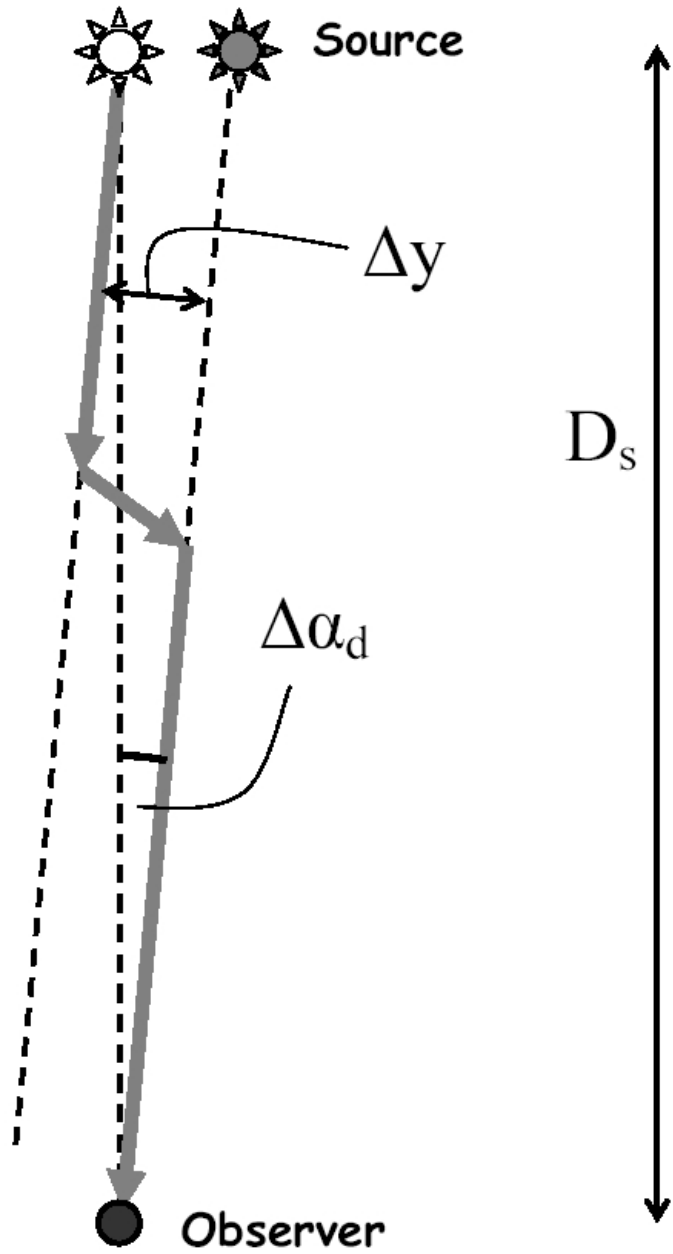}}}
\caption{The observational effect of the displacement in
trajectory of the photon.}
\end{figure}

\end{document}